\documentclass[aps,prl,showpacs,twocolumn,superscriptaddress]{revtex4}
\usepackage{amsmath}
\usepackage{amssymb}
\usepackage{epsfig}
\usepackage{graphicx,amsmath}

\begin{document}
\title{Universal Behavior of Two-Dimensional $\rm ^3He$ at
Low Temperatures}
\author{V.R. Shaginyan}\email{vrshag@thd.pnpi.spb.ru}
\affiliation{Petersburg Nuclear Physics Institute, RAS, Gatchina,
188300, Russia} \affiliation{CTSPS, Clark Atlanta University,
Atlanta, Georgia 30314, USA}
\author{A.Z. Msezane}
\affiliation{CTSPS, Clark Atlanta University, Atlanta, Georgia
30314, USA}
\author{K.G. Popov}
\affiliation{Komi Science Center, Ural Division, RAS, 3a, Chernova
street Syktyvkar, 167982, Russia}
\author{V.A. Stephanovich}\email{stef@math.uni.opole.pl}
\affiliation{Opole University, Institute of Mathematics and
Informatics, Opole, 45-052, Poland}

\begin{abstract}
On the example of two-dimensional (2D) $\rm ^3He$ we demonstrate
that the main universal features of its experimental temperature $T$
- density $x$ phase diagram [see M. Neumann, J. Ny\'{e}ki, J.
Saunders, Science {\bf 317}, 1356 (2007)] look like those in the
heavy-fermion metals. Our comprehensive theoretical analysis of
experimental situation in 2D $\rm ^3He$ allows us to propose a
simple expression for effective mass $M^*(T,x)$, describing all
diverse experimental facts in 2D $\rm ^3He$ in unified manner and
demonstrating that the universal behavior of $M^*(T,x)$ coincides
with that observed in HF metals.
\end{abstract}
\maketitle

One of the main purposes of condensed matter physics is to unveil
the nature of the non-Fermi liquid (NFL) behavior in various
strongly correlated Fermi-systems. These substances, such as high
temperature superconductors, heavy fermion (HF) metals, 2D electron
and $\rm ^3He$ systems are the objects of intensive studies leading
to understanding of many-body effects and quantum phase transitions
responsible for the NFL behavior. Heavy fermion metals provide
important examples of strongly correlated Fermi-systems
\cite{Stew2001,loh}. In these compounds, being f-electron alloys, a
lattice of f-electron spins couples to the itinerant electronic
system by $s-f$ Kondo exchange interaction. The properties of such
systems are now hotly debated as there is a common wisdom that they
are related to zero temperature quantum fluctuations, suppressing
quasiparticles and giving rise to a quantum critical point (QCP),
where the systems transit to different magnetic ground states
generating their specific NFL behavior \cite{loh,col}. On the other
hand, it was shown that the electronic system of HF metals
demonstrates the universal low-temperature behavior irrespectively
of their magnetic ground state \cite{epl2}. Therefore it is of
crucial importance to check whether this behavior can be observed in
2D Fermi systems. Fortunately, the recent measurements on 2D $\rm
^3He$ become available \cite{he3,he3a}. Their results are extremely
significant as they allow to check the presence of the universal
behavior in the system formed by $\rm ^3He$ atoms which are
essentially different from electrons. Namely, the neutral atoms of
2D $\rm ^3He$ are fermions with spin $S=1/2$ and they interact with
each other by van-der-Waals forces with strong hardcore repulsion
(due to electrostatic repulsion of protons) and a weakly attractive
tail. The different character of interparticle interaction along
with the fact, that a mass of He atom is 3 orders of magnitude
larger then that of an electron, makes $\rm ^3He$ to have
drastically different microscopic properties then that of 3D HF
metals. Because of this difference nobody can be sure that the
macroscopic physical properties of both above fermionic systems will
be more or less similar to each other.

The bulk liquid $\rm ^3He$ is historically the first object, to
which a Landau Fermi-liquid (LFL) theory had been applied
\cite{land56}. This substance, being intrinsically isotropic
Fermi-liquid with negligible spin-orbit interaction is an ideal
object to test the LFL theory. Recently 2D $\rm ^3He$ sample has
been fabricated and its thermodynamic properties have been
thoroughly investigated \cite{he3,he3a}. Our analysis of the
experimental measurements has shown that the behavior of 2D $^3$He
is pretty similar to that of 3D HF compounds with various ground
state magnetic properties. Because of van-der-Waals character of
interparticle interaction, $^3$He has a very important feature: the
change of total density of $^3$He film drives it towards QCP at
which the quasiparticle effective mass $M^*$ diverges
\cite{he3,prlhe}. This peculiarity permits to plot the experimental
temperature-density phase diagram, which in turn can be directly
compared with theoretical predictions.

In this letter we show that despite of very different microscopic
nature of 2D $^3$He and 3D HF metals, their main universal features
are the same, being dictated by a Landau quasiparticles paradigm.
Namely, we demonstrate that the main universal features of $\rm
^3He$ experimental $T$ - $x$ phase diagram look like those in HF
metals and can be well captured utilizing our notion of fermion
condensation quantum phase transition (FCQPT)
\cite{xoshag,obz1,amshag,volovik,obz} based on the quasiparticles
paradigm and thus deriving NFL properties of above systems from
modified LFL theory. The modification is that in contrast to the
Landau quasiparticle effective mass, the $\rm ^3He$ effective mass
$M^*(T,x)$ becomes temperature and density dependent. We demonstrate
that the universal behavior of $M^*(T,x)$ coincides with that
observed in HF metals.

Let us consider HF liquid at $T=0$ characterized by the effective
mass $M^*$. Upon applying the well-known equation, we can relate
$M^*$ to the bare electron mass $M$ \cite{land56,pfit}
$M^*/M=1/(1-N_0F^1(p_F,p_F)/3)$. Here $N_0$ is the density of states
of a free electron gas, $p_F$ is Fermi momentum, and $F^1(p_F,p_F)$
is the $p$-wave component of Landau interaction amplitude $F$. Since
LFL theory implies the number density in the form $x =p_F^3/3\pi^2$,
we can rewrite the amplitude as $F^1(p_F,p_F)=F^1(x)$. When at some
critical point $x=x_c$, $F^1(x)$ achieves certain threshold value,
the denominator tends to zero so that the effective mass diverges at
$T=0$ and the system undergoes the fermion condensation quantum
phase transition (FCQPT) \cite{xoshag,obz1,amshag,volovik,obz}. The
leading term of this divergence reads
\begin{equation}\label{zui2}
\frac{M^*(x)}{M}=A+\frac{B}{1-z},\ z=\frac {x}{x_c},
\end{equation}
where $M$ is the bare mass, Eq. (\ref{zui2}) is valid in both 3D and
2D cases, while the values of factors $A$ and $B$ depend on
dimensionality and inter-particle interaction \cite{obz}. At $x>x_c$
the fermion condensation takes place. Here we confine ourselves to
the case $x<x_c$.

When the system approaches FCQPT, the dependence $M^*(T,x)$ is
governed by Landau equation \cite{land56,obz}
\begin{equation}
\frac{1}{M^*(T,x)}=\frac{1}{M}+\int\frac{{\bf p}_F{\bf p}}{p_F^3}F
({\bf p_F},{\bf p})\frac{\partial n({\bf
p},T,x)}{\partial{p}}\frac{d{\bf p}}{(2\pi)^3},\label{LQ}
\end{equation}
where $n({\bf p},T,x)$ is the distribution function of
quasiparticles. The approximate solution of this equation is of the
form \cite{epl2}
\begin{eqnarray}
\frac{M}{{M^* (T)}}&=&\frac{M}{{M^*(x)}}+\beta f(0)\ln\left\{
{1+\exp(-1/\beta)}\right\}\nonumber \\
&+&\lambda _1\beta^2+\lambda_2 \beta^4 + ...,\label{zui1}
\end{eqnarray}
where $\lambda_1>0$ and $\lambda_2<0$ are constants of order unity,
$\beta=TM^*(T)/p_F^2$ and $f(0)\sim F^1(x_c)$. It follows from Eq.
(\ref{zui1}) that the effective mass $M^*(T)$ as a function of $T$
and $x$ reveals three different regimes at growing temperature. At
the lowest temperatures we have LFL regime with $M^*(T)\simeq
M^*(x)+aT^2$ with $a<0$ since $\lambda_1>0$. This observation
coincides with facts \cite{he3, prlhe}. The effective mass as a
function of $T$ decays up to a minimum and afterward grows, reaching
its maximum $M^*_M(T,x)$ at some temperature $T_{\rm max}(x)$ with
subsequent diminishing as $T^{-2/3}$ \cite{ckz,obz}. Moreover, the
closer is number density $x$ to its threshold value $x_c$, the
higher is the rate of the growth. The peak value $M^*_M$ grows also,
but the maximum temperature $T_{\rm max}$ lowers. Near this
temperature the last "traces" of LFL regime disappear, manifesting
themselves in the divergence of above low-temperature series and
substantial growth of $M^*(x)$. The temperature region beginning
near above minimum and continuing up to $T_{\rm max}(x)$ signifies
the crossover between LFL regime with almost constant effective mass
and NFL behavior, given by $T^{-2/3}$ dependence. Thus the $T_{\rm
max}$ point can be regarded as crossover between LFL and NFL
regimes. The latter regime sets up at $T\leq T_{\rm max}$, when
$M^*(x)\to \infty$, giving rise to $T^{-2/3}$ effective mass decay
\cite{obz,ckz}.
\begin{figure} [! ht]
\begin{center}
\includegraphics [width=0.4\textwidth]{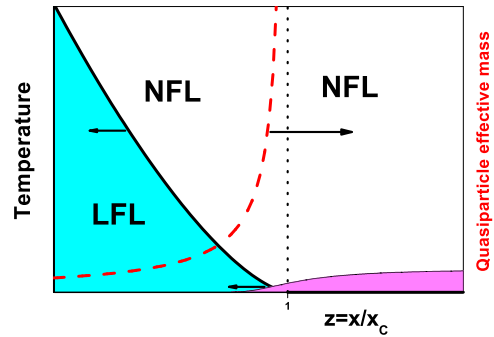}
\end{center}
\vspace*{-1.0cm} \caption{The phase diagram of 2D $^3$He. The part
for $z<1$ corresponds to HF behavior divided to the LFL and NFL
parts by the line $T_{\rm max}(z)\propto (1-z)^{3/2}$, where $T_{\rm
max}$ is the effective mass maximum temperature. The exponent
$3/2=1.5$ coming from Eq. (\ref{zui4}) is in good agreement with the
experimental value $1.7\pm 0.1$ \cite{he3}. The dependence
$M^*(z)\propto (1-z)^{-1}$ is shown by the dashed line. The regime
for $z\geq 1$ consists of LFL piece (the shadowed region, beginning
in the intervening phase $z\leq 1$ \cite{he3}, which is due to the
substrate inhomogeneities, see text) and NFL regime at higher
temperatures.}\label{fd}
\end{figure}

It turns out that $M^*(T,x)$ in the entire $T$ and $x$ range can be
well approximated by a simple universal interpolating function
similar to the case of the application of magnetic field
\cite{obz,epl2,ckz}. The interpolation occurs between LFL
($M^*\propto T^2$) and NFL ($M^*\propto T^{-2/3}$) regimes thus
describing the above crossover. Introducing the dimensionless
variable $y=T/T_{\rm max}$, we obtain the desired expression
\begin{equation}
\frac{M^*(T,x)}{M^*_M} = \frac{M^*(y)}{M^*_M}={M^*_N(y)}\approx
\frac{M^*(x)}{M^*_M}\frac{1+c_1y^2}{1+c_2y^{8/3}}. \label{UN2}
\end{equation}
Here $M^*_N(y)$ is the normalized effective mass,  $c_1$ and $c_2$
are parameters, obtained from the condition of best fit to
experiment. Equation (\ref{zui2}) shows that $M^*_M\propto 1/(1-z)$
and it follows from (\ref{zui1}) that $M^*_M\propto T^{-2/3}$. As a
result, we obtain
\begin{equation}\label{zui4}
T_{\rm max}\propto(1-z)^{3/2}.
\end{equation}
We note that obtained results are in agreement with numerical
calculations \cite{obz,ckz}.
\begin{figure} [! ht]
\begin{center}
\includegraphics [width=0.4\textwidth]{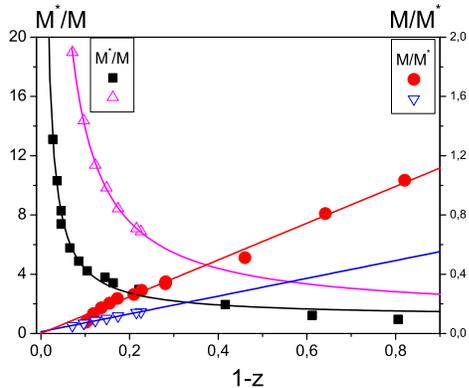}
\end{center}
\vspace*{-1.0cm} \caption{The dependence of the effective mass
$M^*(z)$ on dimensionless density $1-z=1-x/x_c$. Experimental data
from Ref. \cite{prlhe} are shown by circles and squares and those
from Ref. \cite{he3} are shown by triangles. The effective mass is
fitted as $M^*(z)/M\propto A+B/(1-z)$ (see Eq. (\ref{zui2})), while
the reciprocal one as $M/M^*(z)\propto A_1z$, where $A,B$ and $A_1$
are constants.}\label{MXM}
\end{figure}
$M^*(T)$ can be measured in experiments on strongly correlated
Fermi-systems. For example, $M^*(T)\propto C(T)/T\propto
S(T)/T\propto M_0(T)\propto\chi(T)$ where $C(T)$ is the specific
heat, $S(T)$ --- entropy, $M_0(T)$ --- magnetization and $\chi(T)$
--- AC magnetic susceptibility. If the measurements
are performed at fixed $x$ then, as it follows from Eq. (\ref{UN2}),
the effective mass reaches the maximum at $T=T_{\rm max}$. Upon
normalizing both $M^*(T)$ by its peak value at each $x$ and the
temperature by $T_{\rm max}$, we see from Eq. (\ref{UN2}) that all
the curves merge into single one demonstrating a scaling behavior.
\begin{figure} [! ht]
\begin{center}
\includegraphics [width=0.4\textwidth]{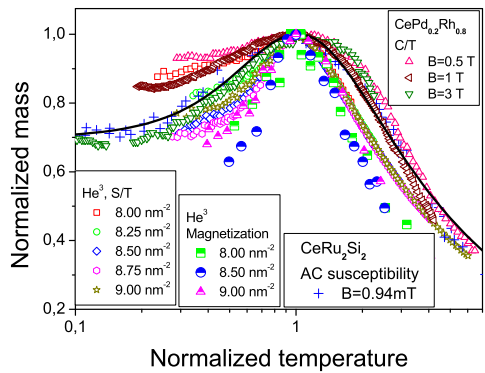}
\end{center}
\vspace*{-1.0cm} \caption{The normalized effective mass $M^*_N$ as a
function of the normalized temperature $T/T_{\rm max}$ at densities
shown in the left down corner. The behavior $M^*_N$ is extracted
from experimental data for $S(T)/T$ in 2D $^3$He \cite{he3a} and 3D
HF
compounds %with different magnetic ground states
such as $\rm{CeRu_2Si_2}$ and $\rm CePd_{1-x}Rh_x$
\cite{pikul,takah}, fitted by the universal function
\eqref{UN2}.}\label{f2}
\end{figure}

In Fig. \ref{fd}, we show the phase diagram of 2D $^3$He in the
variables $T$ - $z$ (see Eq. \eqref{zui2}). For the sake of
comparison the plot of the effective mass versus $z$ is shown by
dashed line. The part of the diagram where $z<1$ corresponds to HF
behavior and consists of LFL and NFL parts, divided by the line
$T_{\rm max}(z)\propto (1-z)^{3/2}$. We pay attention here, that our
exponent $3/2=1.5$ is exact as compared to that from Ref. \cite{he3}
$1.7\pm 0.1$. The agreement between theoretical and experimental
exponents suggests that our FCQPT scenario takes place both in 2D
$^3$He and in HF metals.
\begin{figure} [! ht]
\begin{center}
\includegraphics [width=0.4\textwidth]{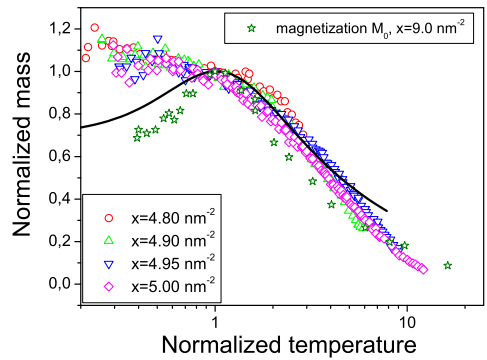}
\end{center}
\vspace*{-1.0cm} \caption{The dependence $M^*_N(T/T_{\rm max})$ at
densities shown in the left down corner. The behavior $M^*_N$ is
extracted from experimental data for $C(T)/T$ in 2D $^3$He
\cite{prlhe} and for the magnetization $M_0$ in 2D $^3$He
\cite{he3}. The solid curve shows the universal function, see the
caption to Fig. \ref{f2}.}\label{PRLC}
\end{figure}
The regime for $z>1$ consists of low-temperature LFL piece,
(shadowed region, beginning in the intervening phase $z\leq 1$
\cite{he3}) and NFL regime at higher temperatures. The former LFL
piece is related to the peculiarities of substrate on which 2D $\rm
^3He$ film is placed. Namely, it is related to weak substrate
heterogeneity (steps and edges on its surface) so that Landau
quasiparticles, being localized (pinned) on it, give rise to LFL
behavior \cite{he3,he3a}. The competition between thermal and
pinning energies returns the system back to FC state and hence
restores the NFL behavior.

In Fig. \ref{MXM}, we report the experimental values of effective
mass $M^*(z)$ obtained by the measurements on $^3$He monolayer
\cite{prlhe}. These measurements, in coincidence with those from
Ref. \cite{he3}, show the divergence of the effective mass at
$x=x_c$. To show, that our FCQPT approach is able to describe the
above data, we present the fit of $M^*(z)$ by the fractional
expression $M^*(z)/M\propto A+B/(1-z)$ and the reciprocal effective
mass by the linear fit $M/M^*(z)\propto A_1z$. We note here, that
the linear fit has been used to describe the experimental data for
bilayer $^3$He \cite{he3} and we use this function here for the sake
of illustration. It is seen from Fig. \ref{MXM} that the data in
Ref. \cite{he3} ($^3$He bilayer) can be equally well approximated by
both linear and fractional functions, while the data in Ref.
\cite{prlhe} cannot. For instance, both fitting functions give for
the critical density in bilayer $x_c\approx 9.8$ nm$^{-2}$, while
for monolayer \cite{prlhe} these values are different - $x_c=5.56$
for linear fit and $x_c=5.15$ for fractional fit. It is seen from
Fig. \ref{MXM}, that linear fit is unable to properly describe the
experiment \cite{prlhe} at small $1-z$ (i.e. near $x=x_c$), while
the fractional fit describes the experiment pretty good. This means
that the more detailed measurements are necessary in the vicinity
$x=x_c$.

\begin{figure} [! ht]
\begin{center}
\includegraphics [width=0.4\textwidth]{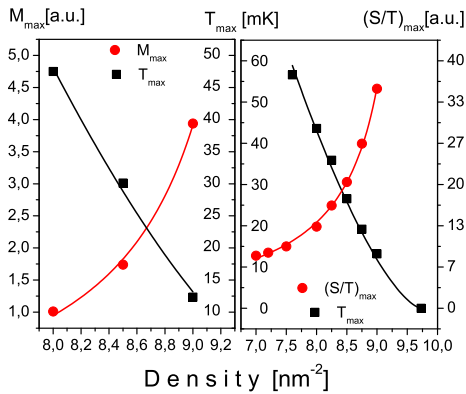}
\end{center}
\vspace*{-1.0cm} \caption{Left panel, the peak temperatures $T_{\rm
max}$ and the peak values $M_{\rm max}$ extracted from measurements
of the magnetization $M_0$ in $^3$He \cite{he3}. Right panel shows
$T_{\rm max}$ and the peak values $(S/T)_{\rm max}$ extracted from
measurements of $S(T)/T$ in $^3$He \cite{he3a}. We approximate
$T_{\rm max}\propto (1-z)^{3/2}$ and $(S/T)_{\rm max}\propto M_{\rm
max}\propto A/(1-z)$.}\label{STM}
\end{figure}

Now we apply the universal dependence \eqref{UN2} to fit the
experiment not only in 2D $^3$He but in 3D HF metals as well.
$M^*_N(y)$ extracted from the entropy measurements on the $^3$He
film \cite{he3a} at different densities $x<x_c$ smaller then the
critical point $x_c=9.9 \pm 0.1$ nm$^{-2}$ is reported in Fig.
\ref{f2}. In the same figure, the data extracted from heat capacity
of ferromagnet CePd$_{0.2}$Rh$_{0.8}$ \cite{pikul} and AC magnetic
susceptibility of paramagnet CeRu$_2$Si$_2$ \cite{takah} are plotted
for different magnetic fields. It is seen that the universal
behavior of the effective mass given by Eq. (\ref{UN2}) (solid curve
in Fig. \ref{f2}) is in accord with experimental facts. All
substances are located at $x<x_c$, where the system progressively
disrupts its LFL behavior at elevated temperatures. In that case the
control parameter, driving the system towards its critical point
$x_c$ is merely a number density $x$. It is seen that the behavior
of the effective mass $M^*_N(y)$, extracted from $S(T)/T$ in 2D
$^3$He (the entropy $S(T)$ is reported in Fig. S8 A of Ref.
\cite{he3a}) looks very much like that in 3D HF compounds. As we
shall see from Fig. \ref{STM}, the amplitude and positions of the
maxima of magnetization $M_0(T)$ and $S(T)/T$ in 2D $^3$He follow
nicely the interpolation formula \eqref{UN2}. We conclude that Eq.
\eqref{UN2} allows us to reduce a 4D function describing the
effective mass to a function of a single variable. Indeed, the
effective mass depends on magnetic field, temperature, number
density and the composition
%of a strongly correlated Fermi-system and
so that all these parameters can be merged in the single variable by
means of interpolating function like Eq. \eqref{UN2}, see also Ref.
\cite{epl2}.

The attempt to fit the available experimental data for $C(T)/T$ in
$\rm ^3He$ \cite{prlhe} by the universal function $M^*_N(y)$ is
reported below in Fig. \ref{PRLC}. Here, the data extracted from
heat capacity $C(T)/T$ for $^3$He monolayer \cite{prlhe} and
magnetization $M_0$ for bilayer \cite{he3}, are reported. It is seen
that the effective mass extracted from these thermodynamic
quantities can be well described by the universal interpolation
formula \eqref{UN2}. We note the qualitative similarity between the
double layer \cite{he3} and monolayer \cite{prlhe} of $^3$He seen
from Fig. \ref{PRLC}.

On the left panel of Fig. \ref{STM}, we show the density dependence
of $T_{\rm max}$, extracted from measurements of the magnetization
$M_0(T)$ on $^3$He bilayer \cite{he3}. The peak temperature is
fitted by Eq. \eqref{zui4}. At the same figure, we have also
reported the maximal magnetization $M_{\rm max}$. It is seen that
$M_{\rm max}$ is well described by the expression $M_{\rm max}
\propto (S/T)_{\rm max}\propto (1-z)^{-1}$, see Eq. (\ref{zui2}).
The right panel of Fig. \ref{STM} reports the peak temperature
$T_{\rm max}$ and the maximal entropy $(S/T)_{\rm max}$ versus the
number density $x$. They are extracted from the measurements of
$S(T)/T$ on $^3$He bilayer \cite{he3a}. The fact that both the left
and right panels have the same behavior shows once more that there
are indeed the quasiparticles, determining the thermodynamic
behavior of 2D $^3$He (and also 3D HF compounds \cite{epl2}) near
their QCP.

To conclude, we have described the diverse experimental facts
related to temperature and number density dependencies of
thermodynamic characteristics of 2D $^3$He by the single universal
function of one argument. The above universal behavior is also
inherent to HF metals with different magnetic ground states. Also,
for the first time, the amplitude and positions of the maxima of the
magnetization $M_0(T)$ and the entropy $S(T)/T$ in 2D $^3$He as the
functions of density have been analyzed on Fig.5. We obtain the
marvelous coincidence with experiment in the framework of our
theory. Moreover, these data could be obtained for $^3$He only and
thus they were inaccessible for analysis in HF metals. This fact
also shows the universality of our approach. Thus we have shown that
bringing the experimental data collected on different strongly
correlated Fermi-systems to the above form immediately reveals their
universal scaling behavior.

This work was supported in part by the RFBR No. 08-02-00038, DOE and
NSF No. DMR-0705328.

\end{document}